\documentclass[conference]{IEEEtran}
\usepackage{graphicx}

\begin{document}

\title{A Consumer Focused Open Data Platform}

\author{}
\author{\IEEEauthorblockN{Cherlton Millette and Patrick Hosein}
\IEEEauthorblockA{Department of Computer Science\\The University of the West Indies\\St. Augustine, Trinidad\\
cherlton.millette@gmail.com, patrick.hosein@sta.uwi.edu}}

\maketitle

\begin{abstract}
Open data has been around for many years but with the advancement of technology and its steady adoption by businesses and governments it promises to create new opportunities for the advancement of society as a whole. Many popular open data platforms have focused on solving the problems faced by the data producer and offers limited support for the data consumers in the way of tools and supported processes for easy access, modeling and mining of data within the various platforms. In this paper we present the specifications for a data consumer oriented platform for open data, the Data-TAP. We demonstrate how this platform is able to (a) meet the goal of leveraging the data resources of existing Open Data platforms, (b) provide a means to standardize and structure data to allow for the merging of related data resources and (c) use a rule-based method for parsing resources. This paper will show how Data-TAP provides an easy to use and understand interface for making open data friendlier for consumers.
\end{abstract}


\section{Introduction}

In the age of Big Data processing, the power of the Open Data resource has yet to be fully harnessed to the betterment of humankind. This has resulted in skeptical views on the usefulness of Open Data. The Open Data resource consists of data anyone can access, use or share. It is a powerful tool that can be used to enable citizens, researchers and academics to develop key resources and make crucial decisions to improve the quality of life for many around the world.

With the recent initiatives of governments to increase the availability of open data to the public, the Open Data resource has steadily grown. However, the approach of many “data publishers” to opening their data is likely to produce a sea of inter-operable but related raw data streams, only decipherable by the technological elite \cite{dawes2012realistic}. This “digital divide” will effectively alienate many “data consumers” who are unable to acquire or employ the technical skills to access or decipher open data.

Over the years, there has been the development of numerous open data platforms that adhere to may of the open data policy guidelines \cite{sunlight}. Unfortunately, many of the platforms developed have focused on solving the problems faced by the data publisher, exasperating the problem outlined. On the other hand, platforms that have attempted to provide solutions that cater to both the data publisher and consumer have seen uphill adoption rates by data publishers as “extra work” is given to data publishers in the form of preparing data for upload. In this paper, we present the Data Transformation and Aggregation Platform (Data-TAP). The Data-TAP is aimed for use by the data consumer and seeks to leverage the data already present in existing data repositories and open data platforms. 

Data-TAP will be the first consumer oriented open data platform that will (1) Leverage the data resources of existing Open Data platforms, (2) Provide a means to standardize and structure data, (3) Use a rule-based method for parsing resources by providing an easy to understand interface for defining parsing rules, (4) Allow for the merging of data resources from differing data repositories, (5) Provide raw API access to data stored on the platform in JSON and CSV formats in alignment with the standards defined for data-sets.

\section{Requirements of an Open Data Platform}

To best serve the needs of the public, open data platforms are required to be highly available services that provide reusable data which is universally available and consumable. In addition, the data published to such platforms must adhere to the following seven principles – Data must be Complete, Primary, Timely, Accessible, Machine readable, Non-discriminatory and Non- Proprietary \cite{tauberer}. A fully fledged open data platform solution that meets all the criteria for providing open data and promotes the seven principle for openness will cater to the needs of both the data publisher and the data consumer.

\subsection{Highly Available Data}

Any system or component that can remain operational and accessible for an indefinite period of time or has sufficient fail over components to avoid any service outages is considered to be Highly Available. When the term high availability (HA) is mentioned, one may create associations with large data centers and with redundant servers and 100\% up-time on services. This concept of always available services is translated into always available data. This means open data platform designers must choose the right technologies that would help their respective open data platforms perform as highly available services.

\subsection{Reusable and Distributive Data}

In a programming context, re-usability is the extent to which code or program assets that can be used in different applications with minimal changes being made to the original code or asset. Distribution of any code or program asset is limited by the legalities involved in product distribution and the rights to the code or asset.

Data can be viewed as a program asset as programs require data to function and be useful. Thus, licensing for parts or whole data-sets will allow for the proper distribution of data that protects the rights of data owners.  It is also noted, re-usability is the ability for data to be accessed for use in many applications of differing functionality with limited changes required to the data. The way to achieve this objective is through data standardization.

\subsection{Universally Consumable}

This concern deals more closely with data quality and ease of use. The data itself must be applicable to real issues and solution formation, and must be easy to digest and understand by users who are not expert in the field of study presented by a given dataset. This would require that data come accompanied with sufficient meta-data tags and descriptions that would help describe the dataset and its contents to non-expert users of the data.

From an application development perspective, this would speak to the need to have all the related data served in a single format along with all the required meta-data needed to understand the purpose of the data. This reduces the complexity of applications' data consumption since all data sources are expected to be delivered in a single, predictable format and helps developers understand how best to use the data.

\section{Review of Available Open Data Platforms}

In this section we review some of the more popular open data solutions \cite{civic} as well as a novel one. Platforms such as CKAN contain features such as Data-Proxy \cite{urbanek} which became the CKAN datastore that provides partial support for Data-TAP features. The review presented in Table \ref{comp} summarizes the differences.

\subsection{CKAN}
CKAN is the most popular Open Data platform in adoption. Developed and maintained by The Open Data Foundation, CKAN’s goal is to provide tools to streamline the publishing, sharing and finding of data. CKAN is aimed at national and regional governments, companies and organizations wanting to make their data open and available.

The platform's mission statement is presented on their website as follows: “{\it CKAN is a powerful data management system that makes data accessible by providing tools to streamline publishing, sharing, finding and using data. CKAN is aimed at data publishers, such as national and regional governments, companies and organizations, wanting to make their data open and available}" \cite{okf}.

The CKAN solution is an open source project built on an open technology stack. It can be installed as an on-premise solution or hosted in the cloud. CKAN is built using python in the back-end, JavaScript in the front-end and stores data using a Postgres database. CKAN uses its internal model to store meta-data about the different records and presents it on a web interface that allows users to browse and search this meta-data. It also offers a powerful API that allows third-party applications and services to be built around it.

\subsection{Socrata}

Socrata is a cloud-based SaaS (Software as a Service) open data catalog platform that provides API, catalog, and data manipulation tools. On its website, Socrata is described as:
“{\it Cloud-based solution for federal agencies and governments of all sizes to transform their data into this kind of vital and vibrant asset. It is the only solution that is dedicated to meeting the needs of data publishers as well as data consumers}" \cite{socrata}. This platform presents one of the few attempts at a holistic approach in building Open Data solutions for both the data publisher and data consumer.

One distinguishing feature of Socrata is that it allows users to create views and visualizations based on published data and save those for others to use. This feature allows for data creation from within the platform itself. Additionally, Socrata offers an open source version of their API that is intended to facilitate transitions for customers that decide to migrate away from the SaaS model. 

\subsection{Junar}

Junar is another cloud-based SaaS open data platform so data is typically managed within Junar’s infrastructure (the “all-in-one” model). Junar can either provide a complete data catalog or can provide data via an API to a separate user catalog. It has some of the most advanced publishing features as it allows users to scrape data from almost any source (web pages, spread sheets, databases and other file formats).
On their website it is described as: “{\it Junar provides everything you need to open data with confidence. It’s built for massive scale and supports local and global data-driven organizations. With Junar, you can easily choose what data to collect, how you want to present it, and when it should be made publicly available. You can also determine which data-sets are made available to the public and which data-sets are available only for your internal use. Think of it as a next generation data management system for sharing information}" \cite{junar}.

Junar is especially aimed at governments and national institutions. It has very close comparisons to the CKAN platform with more polished features available. It also incorporates various analytic and usage tracing features, leveraging trusted platforms such as Google Analytic.

\subsection{Real-time Open Data Platform (RTOD)}

The previous platforms are meant for static data-sets. However with the realization of the Internet of Things (IoT) there is expected to be a deluge of real-time or near real-time data sources. In order to collect and process this data for use by real-time applications the Real-Time Open Data Platform or RTOD was developed \cite{lutchman2015open}. 

With this platform one firsts registers a real-time data stream so that the platform can allocate the required resources. Once streaming of the data begins data consumers can then subscribe for the raw stream or some processed version of the stream. Data consumers can therefore subscribe to multiple data streams and combine this information in novel ways for use in their web or mobile applications. With the advent of IoT this platform can serve as a basis for data collection and dissemination for Smart Grids, Smart Cities and other sensor based networks. The platform also has some basic visualization features so that data consumers can sample the data to see if it is appropriate for their particular needs.

\subsection{Platform Comparison}

In Table \ref{comp} we provide a summary of the differences among the various platforms so far described. RTOD is not included since it's objective is slightly different to the others (i.e., real-time data-sets).

\begin{table*}[!t]
\caption{Comparison of Open Data Platforms}
\label{comp}
\begin{center}
%\bgroup
\def\arraystretch{2} 
\begin{tabular}{|p{2.5cm}|p{4cm}|p{3.2cm}|p{3.2cm}|p{3.2cm}|}
\hline
\textbf{Platform} & \textbf{CKAN} & \textbf{Socrata} & \textbf{Junar} & \textbf{Data-TAP} \\ \hline
\textbf{Target Audience} & Data Publishers & Data Publishers and Consumers & Data Publishers & Data Consumers \\ 
\hline
\textbf{Software License} & Open Source & Proprietary & Proprietary & Open Source \\ \hline
\textbf{Features} & Web based GUI, User group and organization management, Management of Resources and Datasets, Insert and search via Keywords, tags, comments etc., RESTful APIs for all GUI functions, Data Preview and visualizations, User security and authentication mechanism, Linked Data and RDF, Multilanguage, Statistics and Usage tracking. & Web based GUI, Key word search, Meta-data, Data import and exports, Data visualizations and previews, Multilanguage, Statistics and Usage tracking, User management and security & Web based GUI, Key word search, Meta data, Data import and exports, Data visualizations and previews, Multilanguage, Statistics and Usage tracking, User management and security & Web based GUI, Meta-data, Data import and exports, User management and security \\ \hline
\textbf{Data Transformation} & No & Partial -- Supports tool to create new data from existing data. & Partial &  Yes\\ \hline
\textbf{Data Aggregation} & Manual & Manual & Manual & Yes \\ \hline
\textbf{Web API} & Yes - Access to CKAN DataStore which returns meta-data on sets and resources. & Yes & Yes & Yes -- Allows for advanced querying of the DataStore. \\
\hline
\end{tabular}
%\egroup
\end{center}
\end{table*}

\section{The Proposed Open Data Platform}

\begin{figure}
\centering
\includegraphics[width=\columnwidth]{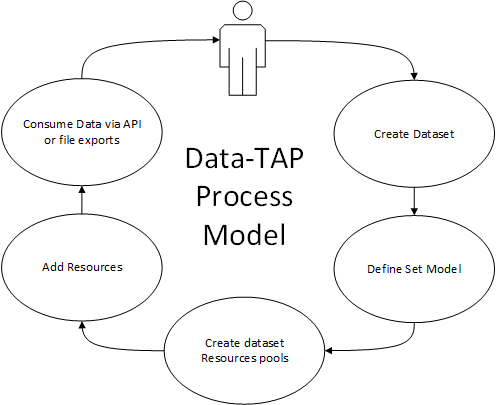}
\caption{A high level overview of the Data-TAP process model}
\label{model}
\end{figure}

In this section we provide details of the proposed open data platform, and in particular, its unique features. A high level model is depicted in Figure \ref{model}.

The Data Transformation and Aggregation Platform is aimed at the data consumer and seeks to leverage the data already present in existing data repositories. In researching the needs of the data consumer we discovered that data consumers fall into three basic categories, the Expert Consumer, the Average Consumer and the Application Developer.
For the scope of this paper we aim to meet the requirements of the Expert consumer and the Application Developer and discuss how existing data repositories can be leveraged.

\subsection{The Data Expert}
The data expert user holds specialized knowledge of the data and its uses in the specified data domain. As such, the expert user is someone who can take the raw data present in data resources and turn them into meaningful data that can be used to tell some story. These stories can be shared via tables of statistics that are then visualized. Data experts may not be the most technologically literate. They may not know how to take full advantage of programming APIs as they may be unable to code or build applications and must rely on Application Developers to fill this gap.

\subsection{The Application Developer}
In contrast to the data expert, we identify the application developer as an expert computer user capable of utilizing well defined APIs to produce useful programs. The application developer's expert knowledge of the data in the domain of interest may not be present, hence their ability to properly build useful data-sets for applications may be hindered. Currently only a hybrid of these users (a single person processing both skills or a team containing persons with each skill) is able to fully utilize available open data resources.

\subsection{Platform Overview}
The Data Transformation and Aggregation platform aims to create a process that allows users to define and redefine data from Open Data resources with limited to no programming skills required. Such a platform will greatly increase the consumption of related Open Data resources and allow the cross examination of data-sets through the use of multiple independent sources. Sets defined in this way will be made available for any consumer to use as-is or, if further processing is required, repeat the process creating new views and perspectives of Open Data. These sets will contain enough meta-data to allow any non-expert user access to the deeper meanings within the data.

Figure \ref{model} gives a high level overview of the Data-TAP process model. When a user creates a new data-set they must define a set model or schema. The schema is a representation containing the data attributes and meta-data that will be made available via the API. The model contains meta-data for each attribute to help describe its usage. The user then creates a resource pool where resources are added either by directly uploading files or linking resources from online sources. 

On creating the resource pool, the user must define rules that will dictate how to interpret resources added to the pool. These rules will transform the imported resources to match the data-set API schema previously defined. Finally, the data can be consumed via the API or through exports to file formats such as CSV.
This process model gives complete control to the data consumer. The data consumer can now define what they want to see up front and have resource contributions fit into this model. This allows for the platform to reach its objective to standardize data within data-sets. It also eliminates any data standardization work to be done by the developers as they are guaranteed the API and data format will not change. This is a huge improvement from the CKAN model which guarantees a stable API but non-standard raw data.

\section{Platform Features}

Now that we have an understanding of the processes that will be carried out by the Data-TAP we can begin to look at what features need to be implemented. To achieve the process work-flow previously defined, the following core user features were developed.

\subsection{User registration} 
The registration process will only apply to users who plan to contribute by creating data-sets and uploading or linking resources. Accessed data should remain free in accordance with open data principles. No matter the dataset, all data will be open and freely available via the API. The web portal will also be able to list and show all data-sets and resources available and, when logged in, a user will be able to view and manage data-sets created by them.

\subsection{Data-set} 
In the Data-TAP model, a data-set refers to the collection of data items extracted from the resources within the associated resource pools. In addition to the collection, the data-set object contains a schema to which each data item must conform to and meta-data that describes the collection as a whole. Only registered users are allowed to create data-sets. During the creation process the user is first required to provide the meta-data to describe the set to other viewers. The next step is to create the set schema used to define the structure of data items within the set. The user will be required to provide the names of each schema attribute, a description and an expected data type for the data values to be read as.

\subsection{Resource Pools} 
To make data-sets useful, the user must create resource pools to add resources. Resources will be either linked from external web sources and other Open Data platforms such as CKAN or directly imported via file uploads. A sample of the resource will be provided for the user to define parsing rules for the pool. Once the rules are created the resources will be parsed and imported.

\subsection{The Data API}
The data API is designed to be simple and concise. It exposes a bare-bones implementation for querying the underlying database. This approach will allow API users the ability to access and find data using the tools available within the underlying database. It will help to simplify the development process. The API has a read-only implementation to help protect data.
The features described are made available through an easy to use and navigate web interface. Having a web-based application allows for cross platform compatibility. Any device with a web browser will be able to access the platform and perform various functions. In addition to these user features, the platform supports specialized background processing features. These background processing features are as follows:
\paragraph{Application Core} This is the main system module. The application core orchestrates all operations within the platform. It is directly responsible for application security, database access and module management.
\paragraph{Rule Generation module} The rule generating engine is responsible for converting user input into parsing rules. The rule generation is built as a snap-in module to allow for the flexibility of adding new rule types which may require new rule generation algorithms to be created.
\paragraph{Transformation and Aggregation Engine} The engine is influenced by the parsing rules generated on the platform. Thus, the Transformation and Aggregation engine is built using a modular approach. A standard interface is provided for passing rules and accessing transformation objects. \cite{bellahsene2011schema}
\paragraph{The API Layer} The API is the forward facing interface that interacts with the front-end web application. It is also the point of entry for any developers wanting to query the dataset available on the platform.

\section{Platform Implementation}
\begin{figure}[!t]
\includegraphics[width=\columnwidth]{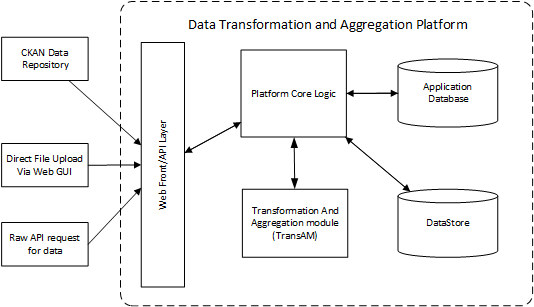}
\caption{The system architecture of the Data-TAP.}
\label{system}
\end{figure}

In this section we will be reviewing the implementation as it pertains to the current prototype that has been built. The platform architecture design shown in figure \ref{system} gives a high level view of the system components. These components can be classified into two broad groups. These classes are the platform modules and the persistence layer. We fist look at the persistence layer.

\subsection{Persistence Layer}
The database design is one of the most important design elements to be considered when developing any application. The Data-TAP makes use of two databases namely the Relational database Postgres SQL and the NoSQL MongoDB database solution. The Postgres database is responsible for serving application data. These objects are modeled within the application as Data-sets, Resources, Resource Pools, Schemas and Users.

The application DataStore is served by MongoDB. The choice of MongoDB over the already implemented Postgres database comes down to schema flexibility. MongoDB allows greater flexibility in storing data when the schema is not known upfront. Thus, for the DataStore which will hold the user defined dataset model structures and the rules for parsing, MongoDB is a perfect fit. 

The philosophy used in choosing the database solutions was not a one size fits all approach. Instead, careful consideration of what is needed to be stored and which database tool best fits the data model and intended use led to this hybrid database approach. 
Highly relational data, such as the application data, is best served in a RDBMS as they have all the bells and whistles to ensure data integrity and correctness via the strict ACID transactional data model. The BASE model present in MongoDB is much better suited for the ad hoc nature of the transformed data models defined at run-time by the user.

\subsection{Platform Modules}
The platform technology on which the Data-TAP is built is Node.js serverside JavaScript. Node.js was chosen because it produces very fast web applications and it is easy to develop prototypes with it. Express is a minimal and flexible Node.js web application framework that provides a robust set of features for building scalable web and mobile applications. Express is the web framework used to build the web interfaces and organize the back-end structure of the application.

The front-end  is designed as a hybrid solution relying on serverside page generation and AJAX calls to data APIs. Most of the application front-end is built at run-time on the Express server using the Jade templating language. The Jade templating engine is a powerful HTML generation tool that allows many of the advanced features.
The project is structured as a typical MVC project with models, views and controllers present. Controllers are in the routes folder for this project. The application start point is server.js. Here we load the Express application into memory which in turn loads the routes/controller code which is the core of the application platform. 

\section{Sample Use Cases}
The Data Transformation and Aggregation Platform is designed to merge the data of heterogeneous systems. Data is taken from the resource and transformed to match the dataset schema. With each resource now having a standard schema, the resources of these once disjoint systems, is now aggregated into a single accessible set. The following are a few use cases where such features would be useful.
	
\subsection{Scenario 1}
In this scenario we look at the many government agencies, businesses and organizations that release transport data to Open data platforms or provide the statistical data in Excel and other file formats. On Google Code in the PublicFeeds section one can find hundreds of resources that offer transport information from around the world. In these organizations the data is presented in isolation of each other limiting the scope and usability of this data. The aggregation of this data into a single cohesive data set would give rise to structured Big Data on transportation.

Using the Data-TAP, a community effort can be utilized where users (independent experts, companies and academics) can add resources to the transportation dataset(s). The dataset will define the structure of the final data produced and the required transformation that must be applied to the raw data. This process would involve the creation of multiple resource pools consuming data for each different resource source. These could be web, platform API or uploaded files. 

All data within the newly created dataset(s) is automatically made available as structured data that can be exported for visualization and further processing or accessed directly via the API for reuse in other applications. These set(s) can be used to help promote and improve smart city transportation systems or be mined in big data analysis tools to reveal interesting trends.

\subsection{Scenario 2}
With the increased availability and richness of Open Data resources, application developers often use Open Data resources to fulfill their application data requirements. Data from multiple sources are typically used to produce data mash-ups.

Each data source for the mash-up has its own format and representation. This causes developers to develop specialized data-marts for their applications with specialized stream filters to transform and store data. The Data-TAP can be used to replace these efforts as it provides graphical tools to perform the same types of data manipulations. This will greatly reduce development efforts and allow teams to focus more on the application features than obtaining and managing data. The Data-TAP has the added benefit of allowing this data to be reused and re-purposed in other applications.

\subsection{Scenario 3}
In this final scenario we provide a live case study. On the CKAN platform located at the URL http://data.tt, there exists a dataset populated with data relating to the Trinidad and Tobago Fisheries division. This data comes from the Wholesale Fish Market Reports which are submitted as daily reports. The dataset is maintained by a national government entity which uploads the daily reports in XLS format. This scenario illustrates a real world approach to open data provisioning. 

\begin{figure}[!t]
\includegraphics[width=\columnwidth]{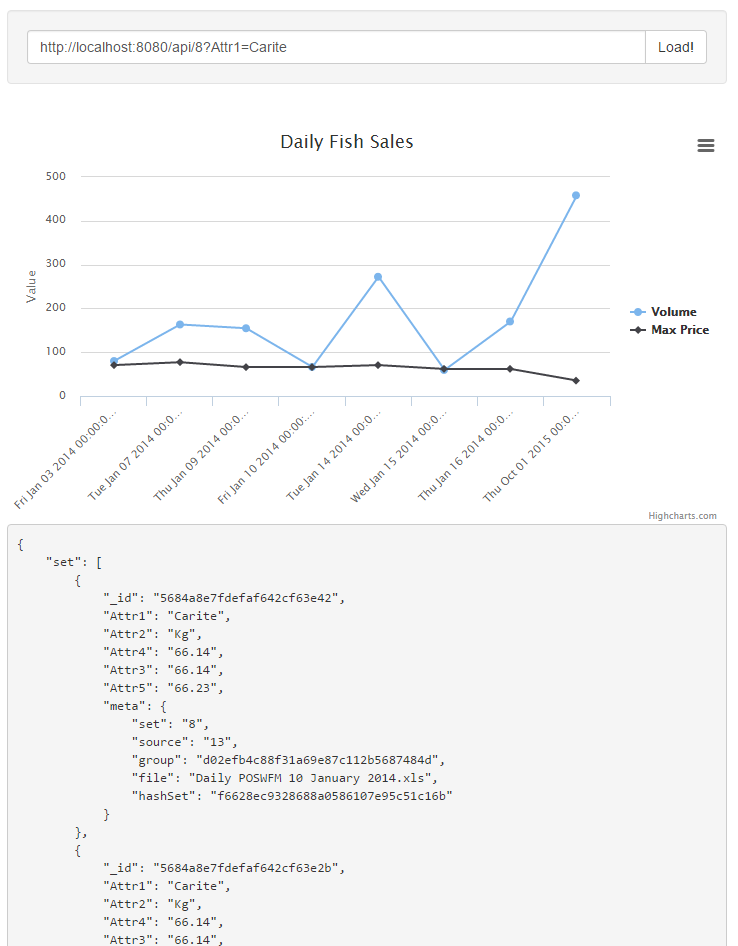}
\caption{Example of Data Aggregation using the platform API.}
\label{app}
\end{figure}

In CKAN any file uploaded with daily information is treated as separate and distinct from any other file containing the same information for a different date. This results in related information being isolated from each other and can create an unclear picture of the information present. To fix this issue, the maintaining authority has the following options (a) The maintainer can choose to create a master file that will contain all the daily data to be presented on CKAN. As daily summaries are generated, the master file is updated with the new details on CKAN. This adds a requirement level to data management processes as the maintainer now maintains the original reports as well as the master file or (b)
The problem can be ignored by the data publisher and so it becomes the duty of the consumer to assemble the data. \newpage In this case the consumer must access each file resource and aggregate them on the fly. The consumer, if savvy enough, can employ the use of databases to cache the data retrieved from the resources.
The second option described is typically the position taken by the data publisher and is the one taken by this publisher as well. There only concern is that the data is made available. It is left up to the consumer to  figure out how best to consume the data.
In this real world scenario, the Data-TAP becomes the data consumers most valued tool as it simplifies and automates the process of merging the resources. The screen capture shown in figure \ref{app} is an application built to show the volume and price trends of a desired produce over a period of time (in this case all time). All the data for this application is retrieved from the Data-TAP APIs. In figure \ref{app} you can see the API URL to the top in the input field. Under the graph you can see the raw data returned by the API.

\section{Conclusions}
In this paper we presented the Data Transformation and Aggregation Platform which is a consumer-focused Open Data platform. We have provided a high level overview of the platform as well as some real world use cases for both expert consumers and application developers 

In the current implementation of this platform the transformation process can only be applied on spread sheet type data. In the future the transformation process will support many more data types such as XML, JSON, geo-spatial data and also support database connectors. In the future, developments to allow deep integration to visualization platforms should be developed to allow for better data consumption.

\bibliographystyle{IEEEtran}
\bibliography{IEEEabrv,od}

\end{document}